\documentclass[12pt]{article}
\usepackage{graphicx}

\begin{document}

\title{Transverse emission of isospin ratios as a probe of high-density symmetry energy in isotopic nuclear reactions}

\author{Zhao-Qing Feng \footnote{Corresponding author. Tel. +86 931 4969215. \newline \emph{E-mail address:} fengzhq@impcas.ac.cn (Z.-Q. Feng)}}
\date{}
\maketitle

\begin{center}
\small \emph{Institute of Modern Physics, Chinese Academy of Sciences, Lanzhou 730000, People's Republic of China}
\end{center}

\textbf{Abstract}
\par
Transverse emission of preequilibrium nucleons, light clusters (complex particles) and charged pions from the isotopic $^{112,124}$Sn+$^{112,124}$Sn reactions at a beam energy of 400\emph{A} MeV, to extract the high-density behavior of nuclear symmetry energy, are investigated within an isospin and momentum dependent transport model. Specifically, the double ratios of neutron/proton, triton/helium-3 and $\pi^{-}/\pi^{+}$ in the squeeze-out domain are analyzed systematically, which have the advantage of reducing the influence of the Coulomb force and less systematic errors. It is found that the transverse momentum distribution of isospin ratios strongly depend on the stiffness of nuclear symmetry energy, which would be a nice observable to extract the high-density symmetry energy. The collision centrality and the mass splitting of neutron and proton in nuclear medium play a significant role on the distribution structure of the ratios, but does not change the influence of symmetry energy on the spectrum.
\newline
\emph{PACS}: 21.65.Ef, 24.10.Lx, 25.75.-q   \\
\emph{Keywords:} transverse emission; isotopic systems; isospin and momentum dependent transport model; high-density symmetry energy

\bigskip

Heavy-ion collisions with neutron-rich beams at intermediate and relativistic energies are a unique tool to extract the information of nuclear equation of state (EoS) of isospin asymmetric nuclear matter under extreme conditions, such as high density, high temperature and large isospin asymmetry etc. The energy per nucleon in the isospin asymmetric nuclear matter is usually expressed as $E(\rho,\delta)=E(\rho,\delta=0)+E_{\textrm{sym}}(\rho)\delta^{2}+\textsc{O}(\delta^{2})$ in terms of baryon density $\rho=\rho_{n}+\rho_{p}$, relative neutron excess $\delta=(\rho_{n}-\rho_{p})/(\rho_{n}+\rho_{p})$, energy per nucleon in a symmetric nuclear matter $E(\rho,\delta=0)$
and nuclear symmetry energy $E_{\textrm{sym}}=\frac{1}{2}\frac{\partial^{2}E(\rho,\delta)}{\partial\delta^{2}}\mid_{\delta=0}$. Typically, two different forms have
been predicted by some microscopical or phenomenological many-body approaches. In one, the symmetry energy increases monotonically with the baryon density, and in the other the symmetry energy increases initially up to a suprasaturation density and then decreases with baryon density. The density dependence of the symmetry energy around and below the normal densities has been roughly constrained and investigated from heavy-ion collisions \cite{Li08,Ba05}. Moreover, the high-density behavior of nuclear symmetry energy predicted by transport models associated with the existing experimental data deviates largely because of the inconsistent treatment of the mean-field potentials and the unclear in-medium nuclear interaction and properties of resonances \cite{Fe06,Xi09,Fe10a,Ru11}, and an opposite conclusion was drawn. Furthermore, new experimental data related to high-density observables and modifications of transport models including the in-medium effects in two-body collisions and in mean-field propagation, are very anticipated. In addition to providing reasonable understanding of the reaction dynamics and the structure of neutron-rich nuclides, the high-density information of symmetry energy also has an important application in astrophysics regarding such subjects as the structure of neutron star, the cooling of protoneutron stars, the nucleosynthesis during supernova explosions of massive stars etc \cite{St05}. The importance of high-density symmetry energy in hadron-quark phase transition
was stated and investigated thoroughly by Di Toro \emph{et al.} \cite{To11}.

The preequilibrium neutron and proton transverse emission in isotopic reaction systems at an incident energy of 50 MeV/A was measured at the National Superconducting Cyclotron Laboratory (Michigan State University, East Lansing, MI, USA) \cite{Fa06} and analyzed with transport models for constraining the symmetry energy at subnuclear densities \cite{Li97,Ts09}, which have been verified as a nice probe for extracting the density dependence of the symmetry energy. The double neutron/proton ratio of free nucleons was investigated in the high-energy range as a probe of symmetry energy at high-baryon density using the isospin Boltzmann-Uehling-Uhlenbeck (IBUU04) model, which was concluded that the symmetry energy results in a turnover spectrum in comparison to the kinetic energy distribution of the lower energy 50 MeV/A \cite{Li06}. In this work, the transverse emission of prompt nucleons and light clusters as well as charged pions produced in the isotopic $^{112,124}$Sn+$^{112,124}$Sn reactions, to extract the symmetry energy at supranormal densities, is investigated within an isospin and momentum dependent transport model (Lanzhou quantum molecular dynamics (LQMD)). The influence of symmetry energy and mass splitting of neutron and proton on the isospin emission is systematically analyzed.

The LQMD model has been updated in order to extract the density dependence of the nuclear symmetry energy from heavy-ion collisions. The main inelastic channels of baryon-baryon and meson-baryon collisions as well as the decay of resonances for producing pion and strangeness particles were implemented in the model \cite{Fe10a,Fe09,Fe10b,Fe10c,Fe11a}. The momentum dependence of the symmetry potential was also included in the model, which results in a splitting of proton and neutron effective mass in nuclear medium \cite{Fe11b}. In the LQMD model, the time evolutions of the baryons (nucleons and resonances ($\Delta$(1232), N*(1440), N*(1535))) and mesons in the system under the self-consistently generated mean-field are governed by Hamilton's equations of motion, which read as
\begin{eqnarray}
\dot{\mathbf{p}}_{i}=-\frac{\partial H}{\partial\mathbf{r}_{i}}, \quad \dot{\mathbf{r}}_{i}=\frac{\partial H}{\partial\mathbf{p}_{i}}.
\end{eqnarray}
Here we only consider the Coulomb interaction for charged hyperons. The Hamiltonian of baryons consists of the relativistic energy, the effective interaction potential and the momentum dependent interaction. The effective interaction potential is composed of the Coulomb interaction and the local interaction
\begin{equation}
U_{int}=U_{Coul}+U_{loc}.
\end{equation}
The Coulomb interaction potential is written as
\begin{equation}
U_{Coul}=\frac{1}{2}\sum_{i,j,j\neq i}\frac{e_{i}e_{j}}{r_{ij}}erf(r_{ij}/\sqrt{4L})
\end{equation}
where the $e_{j}$ is the charged number including protons and charged resonances. The $r_{ij}=|\mathbf{r}_{i}-\mathbf{r}_{j}|$ is the relative distance of two charged particles.

The local interaction potential is derived directly from the Skyrme energy-density functional and expressed as
\begin{equation}
U_{loc}=\int V_{loc}(\rho(\mathbf{r}))d\mathbf{r}.
\end{equation}
The local potential energy-density functional reads
\begin{eqnarray}
V_{loc}(\rho)=&& \frac{\alpha}{2}\frac{\rho^{2}}{\rho_{0}} + \frac{\beta}{1+\gamma}\frac{\rho^{1+\gamma}}{\rho_{0}^{\gamma}} + g_{\tau}\rho^{8/3}/\rho_{0}^{5/3} + E_{sym}^{loc}(\rho)\rho\delta^{2} + \frac{g_{sur}}{2\rho_{0}}(\nabla\rho)^{2}  \nonumber \\
&&  + \frac{g_{sur}^{iso}}{2\rho_{0}}[\nabla(\rho_{n}-\rho_{p})]^{2},
\end{eqnarray}
where the $\rho_{n}$, $\rho_{p}$ and $\rho=\rho_{n}+\rho_{p}$ are the neutron, proton and total densities, respectively, and the $\delta=(\rho_{n}-\rho_{p})/(\rho_{n}+\rho_{p})$ is the isospin asymmetry. The coefficients $g_{\tau}$, $g_{sur}$, $g_{sur}^{iso}$ are related to the Skyrme forces and the parameter Sly6 is taken in the calculation \cite{Fe10b}. The bulk parameters $\alpha$, $\beta$ and $\gamma$ are readjusted after inclusion the momentum term in order to reproduce the compression modulus (here, K=230 MeV) and the binding energy of isospin symmetric nuclear matter at saturation density, which are taken as -296.6 MeV, 197 MeV and 1.143, respectively. The $E_{sym}^{loc}$ is the local part of the symmetry energy, which can be adjusted to mimic predictions of the symmetry energy calculated by microscopical or phenomenological many-body theories and has two-type forms as follows:
\begin{equation}
E_{sym}^{loc}(\rho)=\frac{1}{2}C_{sym}(\rho/\rho_{0})^{\gamma_{s}},
\end{equation}
and
\begin{equation}
E_{sym}^{loc}(\rho)=a_{sym}(\rho/\rho_{0})+b_{sym}(\rho/\rho_{0})^{2}.
\end{equation}
The parameters $C_{sym}$, $a_{sym}$ and $b_{sym}$ are taken as the values of 52.5 MeV, 43 MeV, -16.75 MeV and 23.52 MeV, 32.41 MeV, -20.65 MeV corresponding to the mass splittings of $m_{n}^{\ast}>m_{p}^{\ast}$ and $m_{n}^{\ast}<m_{p}^{\ast}$, respectively. The values of $\gamma_{s}$=0.5, 1., 2. correspond to the soft, linear and hard symmetry energy, respectively, and the Eq. (7) gives a supersoft symmetry energy, which cover the largely uncertain of nuclear symmetry energy, particularly at the supra-saturation densities. All cases cross at saturation density with the value of 31.5 MeV.

A Skyrme-type momentum-dependent potential is used in the model as follows:
\begin{eqnarray}
U_{mom}=&& \frac{1}{2\rho_{0}}\sum_{i,j,j\neq i}\sum_{\tau,\tau'}C_{\tau,\tau'}\delta_{\tau,\tau_{i}}\delta_{\tau',\tau_{j}}\int\int\int d\textbf{p}d\textbf{p}'d\textbf{r}  \nonumber \\
&& \times f_{i}(\textbf{r},\textbf{p},t)[\ln(\epsilon(\textbf{p}-\textbf{p}')^{2}+1)]^{2} f_{j}(\textbf{r},\textbf{p}',t).
\end{eqnarray}
Here $C_{\tau,\tau}=C_{mom}(1+x)$, $C_{\tau,\tau'}=C_{mom}(1-x)$ ($\tau\neq\tau'$) and the isospin symbols $\tau$($\tau'$) represent proton or neutron. The sign of $x$ determines different mass splitting of proton and neutron in nuclear medium, e.g. positive signs corresponding to the case of $m^{\ast}_{n}<m^{\ast}_{p}$. The parameters $C_{mom}$ and $\epsilon$ was determined by fitting the real part of optical potential as a function of incident energy from the proton-nucleus elastic scattering data. In the calculation, we take the values of 1.76 MeV, 500 c$^{2}$/GeV$^{2}$ for the $C_{mom}$ and $\epsilon$, respectively, which result in the effective mass $m^{\ast}/m$=0.75 in nuclear medium at saturation density for symmetric nuclear matter. The parameter $x$ is changed as the strength of the mass splitting, and the values of -0.65 and 0.65 are respective to the cases of $m^{\ast}_{n}>m^{\ast}_{p}$ and $m^{\ast}_{n}<m^{\ast}_{p}$. Shown in Fig. 1 is a comparison of different stiffness of nuclear symmetry energy after inclusion of the momentum-dependent interactions. One can see that all cases cross at saturation density with the value of 31.5 MeV. The local part of the symmetry energy can be adjusted to reflect the uncertain behavior of the symmetry energy at sub- and supra-normal densities.

The evolution of mesons (here mainly pions, kaons and etas) is also determined by the Hamiltonian as follows:
\begin{eqnarray}
H_{M}&& = \sum_{i=1}^{N_{M}}\left( V_{i}^{\textrm{Coul}} + \omega(\textbf{p}_{i},\rho_{i}) \right).
\end{eqnarray}
Here the Coulomb interaction is given by
\begin{equation}
V_{i}^{\textrm{Coul}}=\sum_{j=1}^{N_{B}}\frac{e_{i}e_{j}}{r_{ij}},
\end{equation}
where the $N_{M}$ and $N_{B}$ are the total numbers of mesons and baryons including charged resonances. Here, we use the self-energy in vacuum for pions, namely $\omega(\textbf{p}_{i},\rho_{i})=\sqrt{\textbf{p}_{i}^{2}+m_{M}^{2}}$ with the momentum $\textbf{p}_{i}$ and the mass $m_{M}$ of the pions. The in-medium potentials for pion and kaon propagations can be chosen in the model \cite{Fe10b,Fe11b}. We have included the resonances $\Delta$(1232), $N^{\ast}$(1440), $N^{\ast}$(1535) in nucleon-nucleon (NN) collisions and their decaying into pions and etas. Furthermore, strange particles are directly created by inelastic hadron-hadron collisions \cite{Fe10c}. The probability reaching a channel in a NN collision process is determined by its contribution of the channel cross section to the total cross section. The choice of the channel is done randomly by the weight of the probability.

The preequilibrium nucleons in heavy-ion collisions are mostly produced during the compression stage. Therefore, the high-density information of nuclear phase diagram can be extracted from the preequilibrium nucleon emissions, which are constrained with the transverse momentum and longitudinal rapidity distributions of free nucleons in this work. Shown in Fig. 2 is a comparison of the neutron/proton (n/p) yields of free nucleons within the mid-rapidity bin of $|y/y_{proj}|<$0.3 for the cases of hard and supersoft symmetry energies taken from the $^{124}$Sn+$^{124}$Sn reaction at a beam energy of 400 MeV/nucleon with the mass splittings of $m^{\ast}_{n}>m^{\ast}_{p}$ in the left panel and $m^{\ast}_{n}<m^{\ast}_{p}$ in the right window, respectively. One can see that the distribution structure is sensitive to the collision centrality because of the different high-density reaction zone being formed, but weakly depends on the mass splitting. A flat spectrum at high transverse momentum is observed for the mass splitting of $m^{\ast}_{n}<m^{\ast}_{p}$ because the more energetic neutrons are emitted in the case, which is consistent with the stochastic mean-field (SMF) calculations \cite{Gi10}. One expects to see a larger n/p value with a harder symmetry energy owing to a stronger repulsive (attractive) force acting on neutrons (protons) in the mean-field evolutions. It indeed exists in the near-central heavy-ion collisions. In particular, the difference of the hard and supersoft symmetry energies is larger at low transverse momentum. In the semi-central collisions, the influence of the symmetry energy exhibits an opposite distribution at high transverse momentum because of the secondary collisions with participant nucleons. And a flat spectrum appears for the supersoft symmetry energy. The interest findings are different with the fermi-energy heavy collisions \cite{Li06}.

In order to reduce the systematic errors and the influence of Coulomb force, we calculated the transverse momentum distributions of the double ratios of free nucleons of reaction partners in $^{124}$Sn+$^{124}$Sn over $^{112}$Sn+$^{112}$Sn collisions as shown in Fig. 3. One notices that the symmetry energy results in a variation of the value about 5$\%\sim$10$\%$, in particular, at the low transverse momentum in near central collisions. The hard symmetry energy enhances the ratios because of the larger values at supra-saturation densities. The results are consistent with the IBUU04 calculations except for at high transverse momentum \cite{Li06}. The mass splitting slightly affect the distributions. Experimental measurements are very anticipant to constrain the high-density symmetry energy from the preequilibrium nucleon emissions. Again, the influence of the symmetry energy on the double ratios in the case of peripheral collisions is opposite at high transverse momentum. Overall, the ratio increases with the transverse momentum in both collision centralities. Figure 4 is a comparison of the double $^{3}$H/$^{3}$He ratio of light clusters with hard and supersoft symmetry energies for the different mass splittings of $m^{\ast}_{n}>m^{\ast}_{p}$ in the left window and $m^{\ast}_{n}<m^{\ast}_{p}$ in the right window, respectively. The effect of the symmetry energy only appears at high transverse momentum in the semi-central collisions, but with larger error bars.

Production of pions in heavy-ion collisions at near threshold energies is a primary product in nucleon-nucleon inelastic scattering and has been verified as a sensitive probe of nuclear symmetry energy \cite{Fe06,Xi09,Fe10a}, which can be easily detected in experimentally and be mainly created in the domain at supra-saturation densities of compressed nuclear matter. The double $\pi^{-}/\pi^{+}$ ratios of total charged pions produced in the isotopic systems $^{124}$Sn+$^{124}$Sn over $^{112}$Sn+$^{112}$Sn as a function of transverse momentum are calculated as shown in Fig. 5 at the beam energy of 400 MeV/nucleon for near-central collisions with the mass splittings of $m^{\ast}_{n}>m^{\ast}_{p}$ (left panel) and $m^{\ast}_{n}<m^{\ast}_{p}$ (right panel), respectively. One can see that the effect of symmetry energy is obvious at high transverse momentum. The hard symmetry energy leads to the larger values of the double ratios, in particular at high transverse momentum, which results from the fact that a number of $\pi^{-}$ are produced with the hard symmetry energy and pronounced in neutron-rich systems in the model \cite{Fe10a,Fe10b}. The conclusions does not satisfy the results in Ref. \cite{Yo06}, where a softer symmetry energy leads to a larger value of the double ratio for the energetic pions and an opposite distribution appears at low kinetic energies. The pion production is complicated not only involving the nucleon-nucleon inelastic collisions such as the $NN \rightarrow N\Delta$ and the inverse channel $N\Delta \rightarrow NN$, but also being related to the absorbtion by a nucleon. The mass splitting slightly influences the high-momentum distribution, but does not change the effect of symmetry energy.

In summary, within the transport model LQMD we have investigated the transverse emissions of free nucleons, light clusters and charged pions produced in isotopic reaction systems of $^{124}$Sn+$^{124}$Sn and $^{112}$Sn+$^{112}$Sn at the beam energy of 400 MeV/nucleon. The single and double ratios of isospin particles are analyzed with different collision centralities and effective mass splittings. It is found that the single and double neutron/proton ratios of mid-rapidity nucleon emissions and the double $\pi^{-}/\pi^{+}$ yields are sensitive to the density dependence of the symmetry energy, which are useful probes for constraining the high-density behavior of the nuclear symmetry energy. The collision centrality changes the high-momentum distributions of the single and double ratios, in particular, for the free nucleon emissions. However, the mass splitting slightly affects the transverse emission of the isospin ratios.

Fruitful discussions with Prof. M. Di Toro, Prof. H. Lenske, Dr. M. Colonna and Dr. T. Gaitanos are acknowledged. This work was supported by the National Natural Science Foundation of China under Grant Nos 10805061 and 11175218, the Special Foundation of the President Fund and the West Doctoral Project of Chinese Academy of Sciences. The author is grateful to the support of K. C. Wong Education Foundation (KCWEF) and DAAD during his research stay in Justus-Liebig-Universit\"{a}t Giessen, Germany.

\newpage

%%%%%%%%%%%%%%%%%%%%%%%%%%%%%%%%%%%%%%% figure 1 %%%%%%%%%%%%%%%%%%%%%%%%
\begin{figure}
\begin{center}
{\includegraphics*[width=0.8\textwidth]{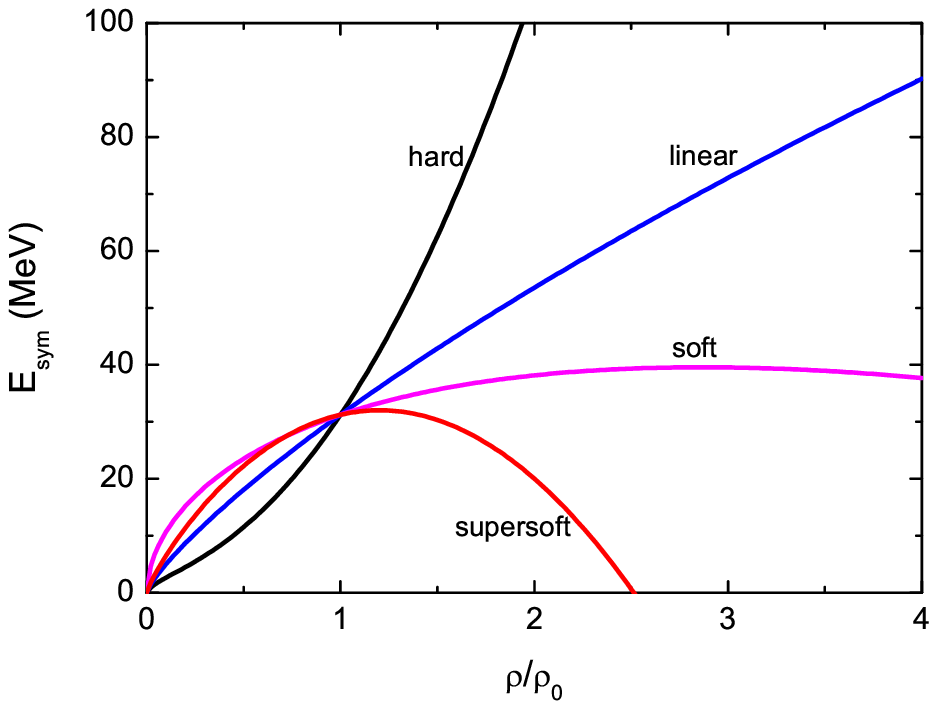}}
\end{center}
\caption{Nuclear symmetry energy as a function of baryon density with the MDI interaction for the different stiffness (supersoft, soft, linear and hard).}
\end{figure}
%%%%%%%%%%%%%%%%%%%%%%%%%%%%%%%%%%%%%%%%%%%%%%%%%%%%%%%%%%%%%%%%%%%%%%%%%

%%%%%%%%%%%%%%%%%%%%%%%%%%%%%%%%%%%%%%% figure 2 %%%%%%%%%%%%%%%%%%%%%%%%
\begin{figure}
\begin{center}
{\includegraphics*[width=1.\textwidth]{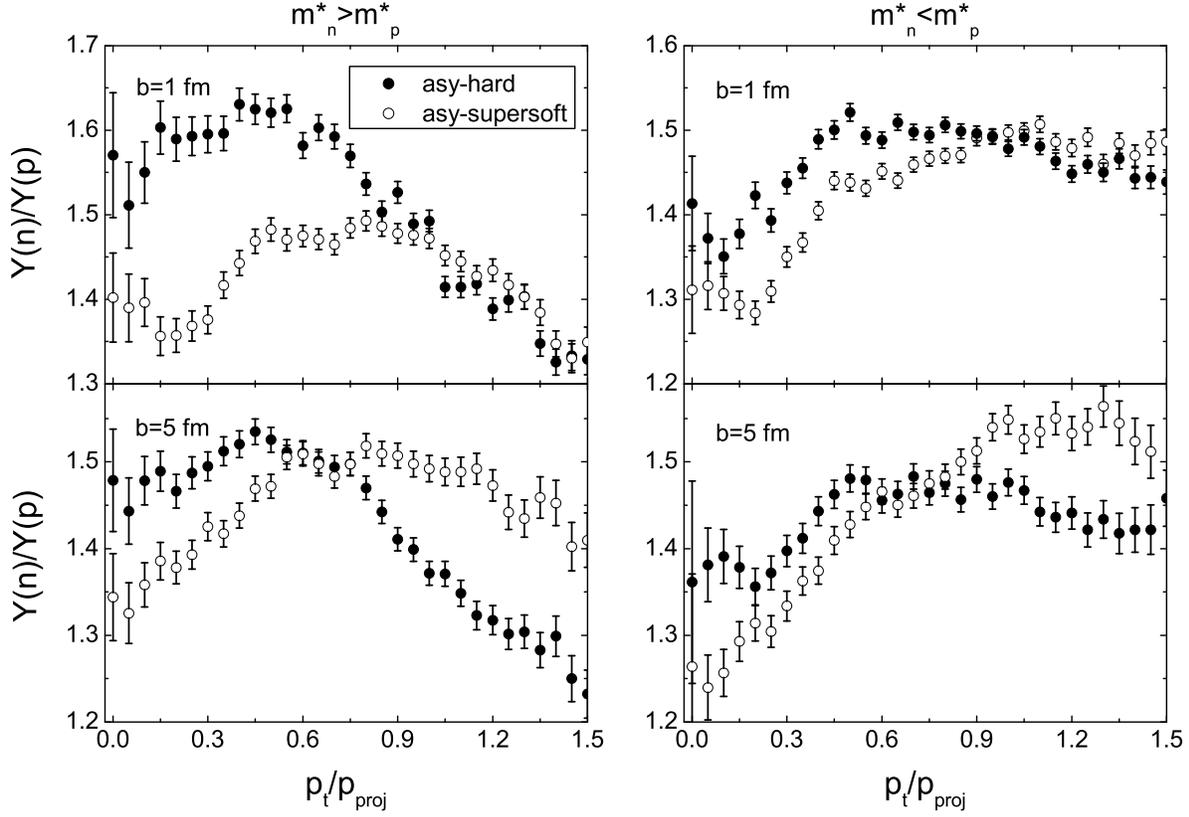}}
\end{center}
\caption{Transverse momentum distributions of neutron/proton ratio of free nucleon emissions within the rapidity bin $|y/y_{proj}|<$0.3 in the $^{124}$Sn+$^{124}$Sn reaction at incident energy of 400 MeV/nucleon for near-central (b=1 fm) and semi-central (b=5 fm) collisions with the mass splittings of $m^{\ast}_{n}>m^{\ast}_{p}$ (left window) and $m^{\ast}_{n}<m^{\ast}_{p}$ (right window), respectively.}
\end{figure}
%%%%%%%%%%%%%%%%%%%%%%%%%%%%%%%%%%%%%%%%%%%%%%%%%%%%%%%%%%%%%%%%%%%%%%%%%

%%%%%%%%%%%%%%%%%%%%%%%%%%%%%%%%%%%%%%% figure 3 %%%%%%%%%%%%%%%%%%%%%%%%
\begin{figure}
\begin{center}
{\includegraphics*[width=1.\textwidth]{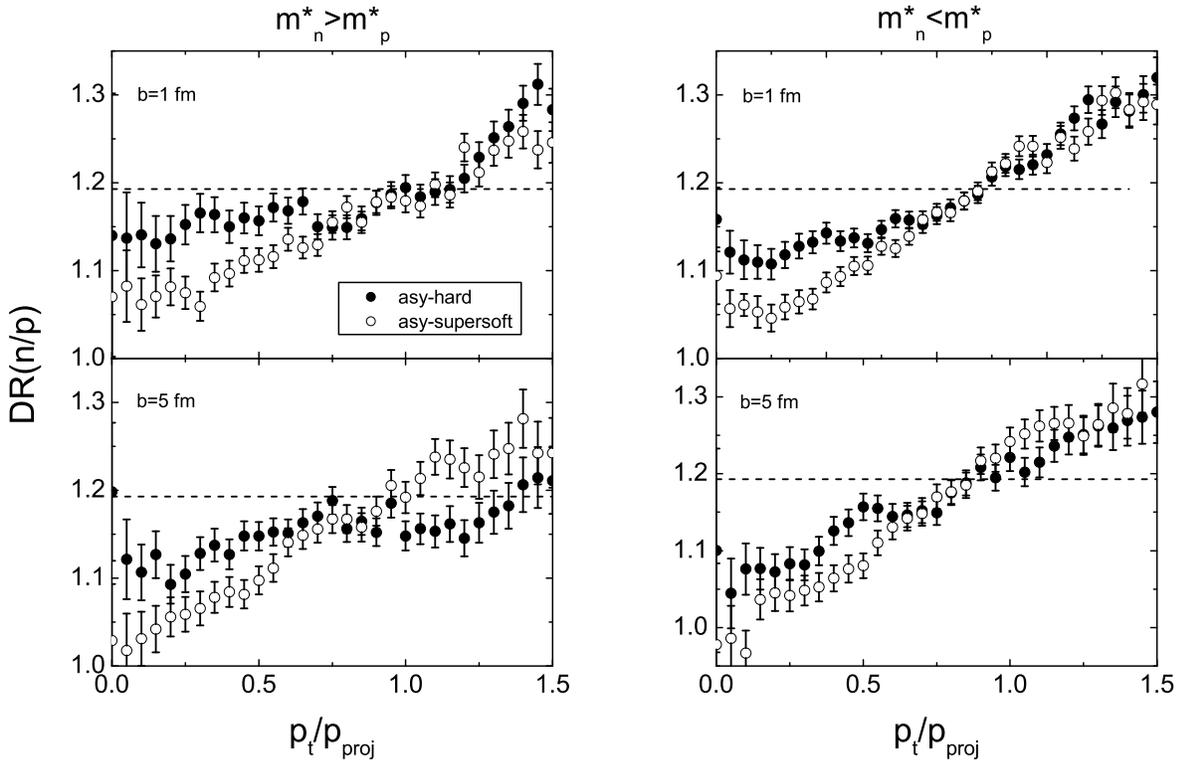}}
\end{center}
\caption{The double neutron/proton ratio of free nucleons taken from the reactions of $^{112}$Sn+$^{112}$Sn and $^{124}$Sn+$^{124}$Sn at an incident energy of 400 MeV/nucleon for different mass splittings of $m^{\ast}_{n}>m^{\ast}_{p}$ (left window) and $m^{\ast}_{n}<m^{\ast}_{p}$ (right window), respectively.}
\end{figure}
%%%%%%%%%%%%%%%%%%%%%%%%%%%%%%%%%%%%%%%%%%%%%%%%%%%%%%%%%%%%%%%%%%%%%%%%%

%%%%%%%%%%%%%%%%%%%%%%%%%%%%%%%%%%%%%%% figure 4 %%%%%%%%%%%%%%%%%%%%%%%%
\begin{figure}
\begin{center}
{\includegraphics*[width=1.\textwidth]{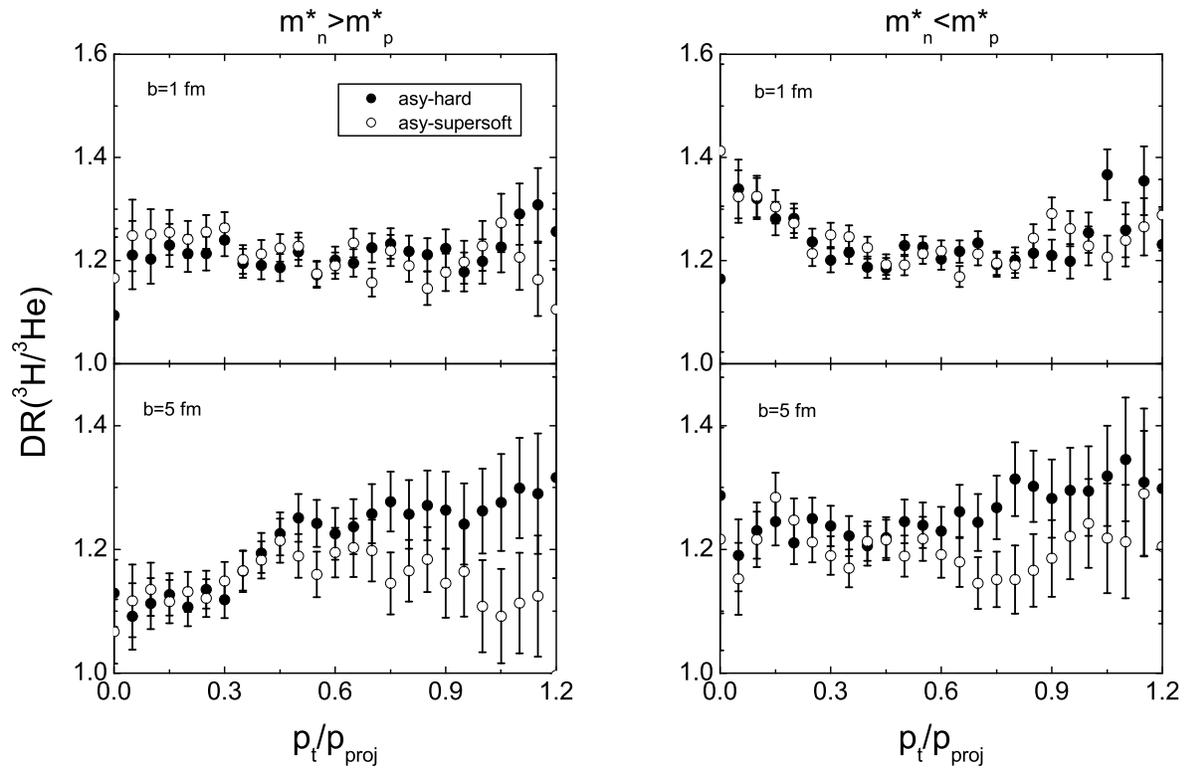}}
\end{center}
\caption{The same as in Fig. 3, but for the triton and helium-3 emissions.}
\end{figure}
%%%%%%%%%%%%%%%%%%%%%%%%%%%%%%%%%%%%%%%%%%%%%%%%%%%%%%%%%%%%%%%%%%%%%%%%%

%%%%%%%%%%%%%%%%%%%%%%%%%%%%%%%%%%%%%%% figure 5 %%%%%%%%%%%%%%%%%%%%%%%%
\begin{figure}
\begin{center}
{\includegraphics*[width=1.\textwidth]{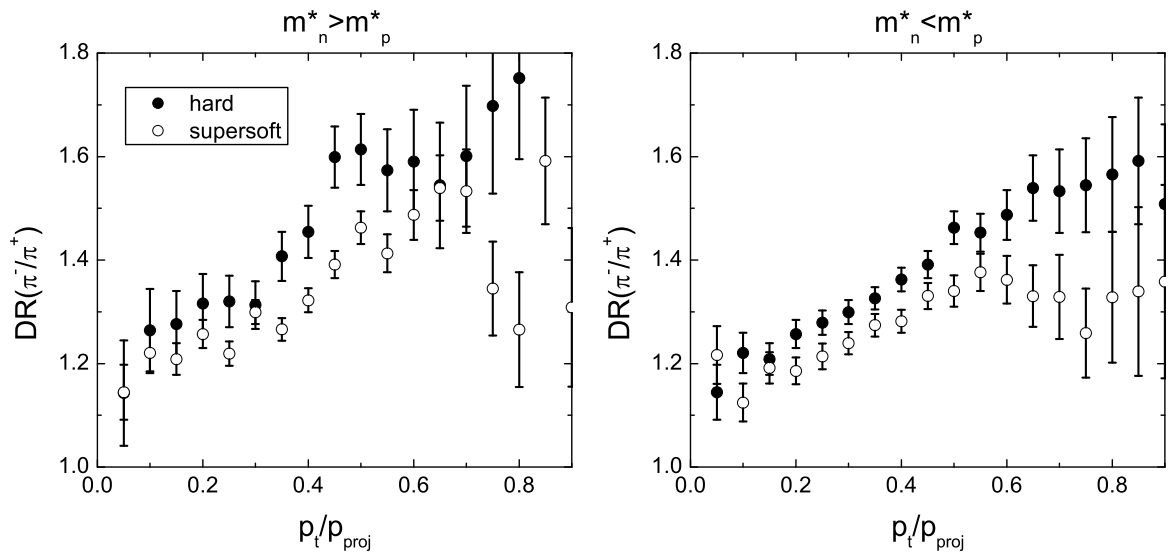}}
\end{center}
\caption{Comparison of the transverse momentum dependence of the double $\pi^{-}/\pi^{+}$ ratio for near-central (b=1 fm) $^{112}$Sn+$^{112}$Sn and $^{124}$Sn+$^{124}$Sn collisions at the incident energy of 400 MeV/nucleon with the hard and supersoft symmetry energy, but for different mass splittings.}
\end{figure}
%%%%%%%%%%%%%%%%%%%%%%%%%%%%%%%%%%%%%%%%%%%%%%%%%%%%%%%%%%%%%%%%%%%%%%%%%

\end{document}